# Self-Organized Nanogratings for Large-Area Surface Plasmon Polariton Excitation and Surface-Enhanced Raman Spectroscopy Sensing

Matteo Barelli[1], Maria Caterina Giordano[1], Pietro Giuseppe Gucciardi[2], Francesco Buatier de Mongeot[1,*]

[1] Dipartimento di Fisica, Università di Genova, Via Dodecaneso 33, I-16146 Genova, Italy

[2] CNR IPCF Istituto per i Processi Chimico-Fisici, viale F. Stagno D'Alcontres 37, I-98156 Messina, Italy

*To whom correspondence may be addressed:* buatier@fisica.unige.it

**TOC GRAPHIC**

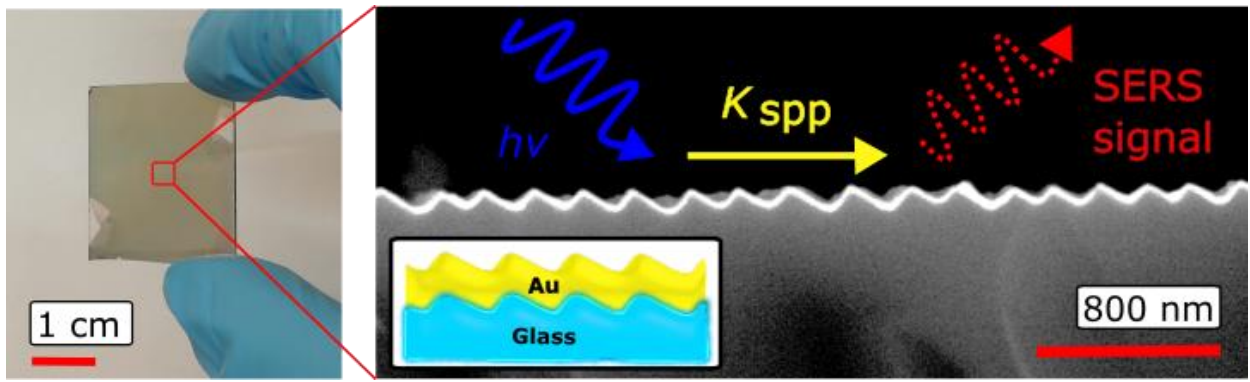

**ABSTRACT**

Surface Plasmon Polaritons (SPP) are exploited due to their intriguing properties for photonic circuits fabrication and miniaturization, for surface enhanced spectroscopies and imaging beyond the diffraction limit. However, the excitation of these plasmonic modes by direct illumination is forbidden by energy/momentum conservation rules. One strategy to overcome this limitation relies on diffraction gratings to match the wavevector of the incoming photons with that of propagating SPP excitations. The main limit of the approaches so far reported in literature is that they rely on highly ordered diffraction gratings fabricated by means of demanding nano-lithographic processes. In this work we demonstrate that an innovative, fully self-organized method based on wrinkling-assisted Ion Beam Sputtering can be exploited to fabricate large area ($cm^2$ scale) nano-rippled soda-lime templates which conformally support ultrathin Au films deposited by physical deposition. The self-organized patterns act as quasi-1D gratings characterized by a remarkably high spatial order which matches properly the transverse photon coherence length. The gratings can thus enable the excitation of hybrid SPP modes confined at the Au/dielectric interfaces, with a resonant wavelength which can be tuned either by modifying the grating period, photon incidence angle or, potentially, the choice of the thin film conductive material. Surface Enhanced Raman Scattering experiments show promising gains in the range of $10^3$ which are competitive, even before a systematic optimization of the sample fabrication parameters, with state-of-the art lithographic systems, demonstrating the potential of such templates for a broad range of optoelectronic applications aiming at plasmon-enhanced photon harvesting for molecular or bio-sensing.

**1. INTRODUCTION**

In the last decades the field of plasmonics witnessed an outstanding development because of the growing interest in the possibility to manipulate optical fields at the nanoscale, well below the diffraction limit, thus boosting a very broad range of optoelectronic applications [1–4]. Different so-called plasmonic modes, with diverse properties and resonant conditions are found in nature [5]. In particular, Surface Plasmons Polaritons (SPP) are electromagnetic modes propagating at the interface between a positive (e.g. a dielectric) and a negative permittivity medium due to resonant oscillations of free carriers [5–8]. Their propagating nature makes them suitable for waveguiding applications, crucial for the development of photonic circuits and components, which are not enabled by Localized Surface Plasmons (LSP) supported by sub-wavelength disconnected nanoparticles [5,6,9–11]. SPP modes can be excited by electro-magnetic radiation under selected wavevector coupling conditions, thus confining the photons energy into subwavelength volumes at the conductive/dielectric interface where the electromagnetic evanescent field is strongly enhanced [5,6]. This peculiar property has recently allowed the highly sensitive near-field detection and imaging of surface plasmon modes with strongly subwavelength spatial resolution at the surface of atomic 2D materials in Scanning Near-field Optical Microscopy (SNOM) [12–15] and molecular nanoimaging in Tip Enhanced Raman Scattering (TERS) [16–19]. In parallel highly sensitive bio-sensing capabilities has been achieved in surface enhanced spectroscopies such as Surface Enhanced Raman Spectroscopy (SERS), Surface Enhanced Coherent Anti-Stokes Raman Spectroscopy (SECARS) and Metal Enhanced Fluorescence (MEF), which provide several orders of magnitude signal gain over their conventional counterparts [4,20–24]. Moreover, travelling SPPs have been exploited to perform Remotely Excited SERS (RE-SERS), a technique where excitation and SERS signal collection are spatially displaced, remarkably reducing the fluorescence background signal originating from diffraction limited excitation volumes [25,26]. Several applications have been recently developed such as optical refractive index sensor devices, already available into the market, which exploit the sensitivity of the SPP resonance to the refractive index of the surrounding dielectric medium [27,28]. Alternatively, the non-radiative plasmon decay generating hot electrons into the conductive material can be exploited in order to enable direct plasmon enhanced photocatalysis [29,30], and/or hot carrier injection into semiconductors, effectively extending the operating range of photonic devices beyond their band-gap limited absorption [31,32]. However, photon-SPP coupling is not allowed by direct illumination because of energy and momentum conservation selection rules, thus particular strategies have to be adopted such as EM coupling via subwavelength probes or diffraction gratings which are able to modulate the wavevector of the incoming photons [5,6]. In particular, diffraction gratings characterized by a very high spatial coherence are typically fabricated via top-down lithographic fabrication methods that, however, can limit the perspective of large-scale, low-cost applications [21–23,33–36].

In this work we demonstrate that a cheap nano-rippled soda-lime glass template, prepared by an innovative, fully self-organized (SO) fabrication method based on wrinkling-enhanced Ion Beam Sputtering [37], can be used as large area support for a thin Au film conformally grown on top of it by thermal vacuum deposition, in a IMI (insulator-metal-insulator) configuration. The resulting sub-wavelength, quasi-1D grating endowed with linearly graded periodicity enables the excitation of a tunable hybrid Surface Plasmon Polariton mode. Based on our knowledge this is the first example of fully self-organized gratings promoting the coherent excitation of SPP modes over large area.

The potential of these large area plasmonic meta-surfaces for highly sensitive bio-sensing have been demonstrated in SERS, showing strong enhancement of the Raman signal promoted by SPP modes. Remarkable SERS gain values in the range of $10^3$, competitive with the figures of lithographic-made systems [38,39] as well as with alternative self-organized approaches [40–43], are found for a pump laser value of 785 nm. Indeed, this makes the sample appealing for molecular sensing, also considering that a systematic optimization of the sample fabrication parameters, which is discussed but it's beyond the scope of this work, could easily further improve these SERS gain values. Finally, we demonstrate how the SERS gain is strongly correlated with the electric field enhancement produced by the excitation of SPP modes by comparing SERS and co-localized micro-extinction measurements. These self-organized gratings represent a very versatile platform, potentially allowing the broadband tuning of the SPP mode from the Near-UV to the Far-IR by simply selecting the conductive layer grown on top of the template, making them interesting for a broad range optoelectronic applications.

## 2. METHODS

2.1 Sample fabrication

A soda-lime glass (2.5×2.5×0.2 cm) is heated up to 680 K and irradiated with a defocused ion beam (Ar$^+$) with a low energy of 800 eV at the incidence angle θ=30° (Fig. 1b) with respect to the sample surface normal direction. Positive charge build-up on the glass surface is prevented during the ion irradiation process by stimulating thermionic electron emission from a tungsten filament negatively biased at $V_{bias}$=-13 V with respect to the glass sample. For ion irradiation prolonged for 1800 s at the pressure of $4 \times 10^{-4}$ mbar (i.e. ion fluence of 1.4x10$^{19}$ ions/cm$^2$), highly ordered faceted nanopatterns evolve on the glass surface showing a steep, asymmetric faceted sawtooth profile. After the nano-rippled pattern is formed on the sample surface, thermal Au deposition is performed on the rippled template at the pressure of about 10$^{-5}$ mbar in a two step process, described in the "Results and discussion" section. The process is designed to grow an Au thin film of approximately homogenous thickness on the slope-modulated glass rippled surface.

2.2 Morphology characterization

The self-organized rippled glass morphology is characterized by means of an atomic force microscope (Nanosurf S Mobile). The average periodicity, line profiles and slope of the nanoripples are extracted by means of WSxM and Gwyddion software from the statistical analysis of AFM measurements [44,45]. Top view back scattered electrons SEM images are acquired by means of a thermionic Hitachi VP-SEM SU3500.

2.3 Optical characterization

Polarized transmission measurements are performed by fiber-coupling the radiation emitted by a compensated deuterium-halogen lamp (DH-2000-BAL, Mikropak) to a linear polarizer, illuminating the sample from the bare glass side. The beam spot diameter in about 1 mm. The signal is then coaxially collected by a high-resolution solid-state spectrometer (HR4000, Ocean Optics). Non-normal incidence transmission measurements are performed by placing the sample on a stage which provides a tilting movement. All the transmission spectra are normalized to the optical transmittance of a bare glass reference substrate tilted at the corresponding sample angle.

2.4 SERS and co-localized extinction measurements

Samples (rippled Au/soda-lime glass and reference flat Au thin film on bare soda-lime glass) are immersed in a $10^{-4}$ M MB solution for an hour and then rinsed in de-ionized water for 10 minutes. Polarized SERS and optical micro-extinction measurements are performed by means of a Horiba XploRA Nano. For SERS, pump lasers at 638 and 785 nm wavelengths are employed, using a 100x objective for the excitation of a diffraction-limited area. Different sample positions, a few mm away from each other, are probed by averaging multiple SERS measurements distanced by some μm. Detection is carried out in backscattering. Micro-extinction measurements have been co-locally performed at corresponding sample coordinates by switching the source to a tungsten lamp illuminating the sample from below and to a 10x objective for signal collection.

## 3. RESULTS AND DISCUSSION

A defocused $Ar^+$ ion beam at low energy (800 eV) irradiates a soda-lime glass sample at an incident angle $\theta$=30° with respect to the surface normal as sketched in Figure 1a-b. The sample temperature is fixed at about 680 K during the $Ar^+$ bombardment (see Methods, paragraph 2.1). The Ion Beam Sputtering process induces a self-organized quasi 1-D nano-rippled morphology all over the glass surface, with propagation vector parallel to the incoming ion beam direction (Fig. 1a). The ripples length extends over several micrometers and the pattern shows a noteworthy degree of long range order. The glass nanostructures are characterized by a high aspect ratio asymmetric sawtooth profile, with vertical dynamic range of about 75 nm (Fig. 1b) and periodicity of approximately 200 nm, assessed by the 2-dimensional self-correlation function of the AFM topography (see Fig. S1). The ripple sides opposed to the incoming ion beam direction develop wider facets with a narrow slope distribution sharply peaked at about +35° (Fig. 1c). Conversely, the sides directly exposed to the $Ar^+$ beam develop narrower facets with a broad slope distribution weakly peaked at about -50° (Fig. 1c). Moreover, we engineered the IBS process in order to produce a sample with a graded periodicity along its length. We achieved this by exploiting an ion dose gradient along the projection of the Ar+ beam, originating from the increasing sample-source distance at the tilted incidence angles here employed, and from the gaussian intensity distribution of the ion beam. Considering a periodicity $\Lambda$=200 nm in the central area of the glass template (Fig. 1a), we achieve in this way a modulation of the grating periodicity of about ±15% moving from the top to the bottom of the sample along the projection of the ion beam (see Fig. S2). It is worth to underline, as described in a recent study by some of the authors [37], that the nanopattern formation kinetics is strongly enhanced by a solid-state wrinkling instability which is activated when the substrate temperature is raised near the glass transition threshold and is concurrently exposed to irradiation with an Ar+ ion beam. Under such conditions a compressive stress builds up in a thin surface skin layer partially depleted of alkali atoms; stress release takes place by surface buckling provided enhanced mass transport in the near-surface region is made possible by the reduction in glass viscosity [37]. We highlight that when ion irradiation is performed at reduced temperatures for which mass transport is strongly reduced, one does not observe the formation of the high aspect ratio faceted wrinkles but rather the low aspect ratio ripples conventionally found after Ion Beam Sputtering of insulating and semiconductor substrates at room temperature [46–48].

As demonstrated by the authors in recent studies [49–51] such faceted, high aspect ratio, ordered rippled patterns are the ideal template for maskless confinement of large area, self-organized arrays of disconnected uniaxial plasmonic nanostructures. The possibility to control tilt and morphology of the nanostructures is an interesting feature for a wide range of applications which exploit tunable dipolar and hybrid localized plasmonic modes. Here instead we propose a novel self-organized fabrication protocol to produce periodically corrugated continuous Au thin films of uniform thickness, conformally supported on the linearly graded glass diffraction grating which sustain propagating Surface Plasmon Polariton (SPP) modes with tunable response.

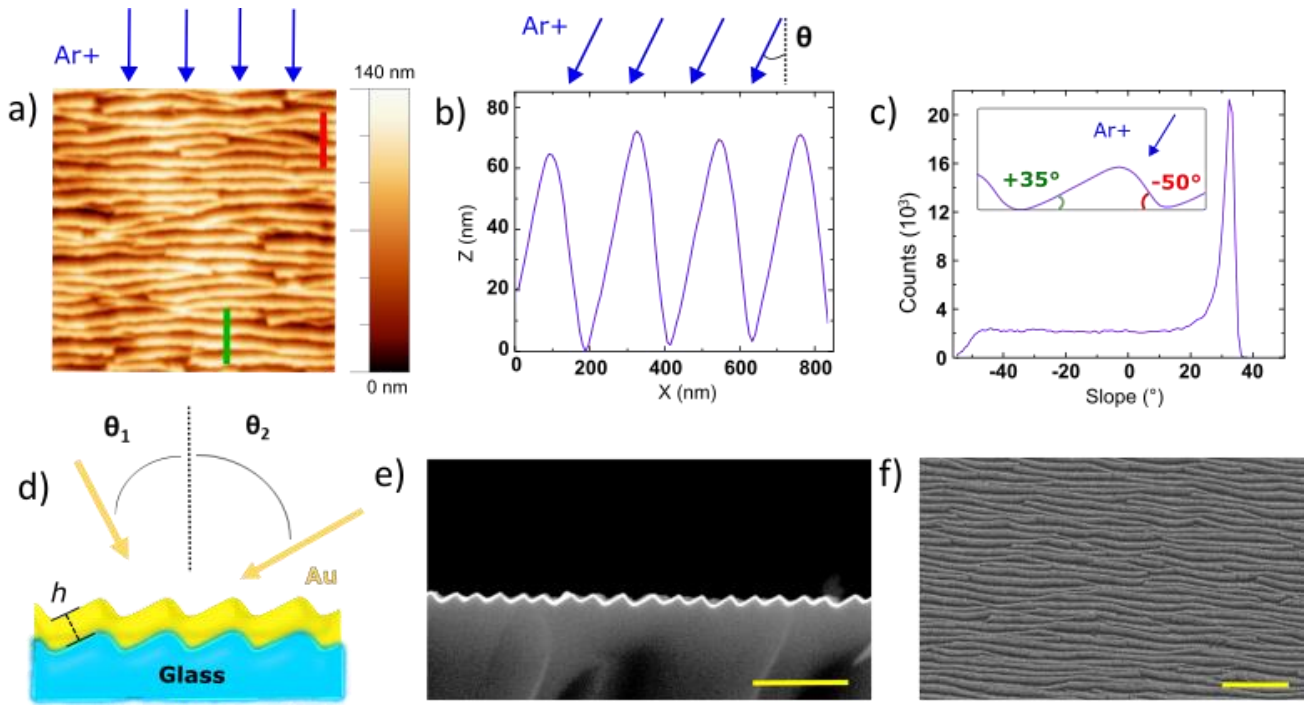

*Figure 1* – a) AFM topography of the rippled glass surface. The red scale bar corresponds to 800 nm. b) AFM line profile corresponding to the green line in panel a). c) Slope distribution histogram extracted by the AFM topography of panel a). d) Cross section sketch of the two step Au deposition process on the asymmetric sawtooth rippled profile of the soda-lime glass surface. e) SEM cross section and f) top view images of the Au covered rippled surface, acquired in the backscattered electrons channel, with acceleration voltage of 15kV. The yellow bars correspond to 800 nm (e) and 2 µm (f).

Given the asymmetric faceted profile of the glass template, a conventional thermal deposition process at normal incidence would result in an Au film with non-uniform thickness, evaluated along the local surface normal. To overcome this limitation, we developed a dual-step off-normal thermal deposition procedure which is capable of achieving growth of an Au film with uniform thickness on the two opposite facets. The first deposition step is performed at the angle $\theta_1$= -15° with respect to the sample surface normal (negative angles corresponding to an anti clock-wise rotation), over the side of the ripple profile opposing the ion beam during the nano-structuring step (Fig. 1d). In this condition the Au beam forms a 20° angle with respect to the local surface normal of the "wide" facets. At the same time an angle of 65° is formed between the Au beam and the local surface normal to the "steep" facets. The Au thickness deposited during the first deposition step on the left side of the ripple can be estimated as $h_{left-1}=h_0 \times \cos 20°$, while on the right side the thickness is $h_{right-1}=h_0 \times \cos 20°$, where $h_0$ is the equivalent Au thickness on a flat substrate. During the experiment $h_0$ is the parameter which is measured, by means of a quartz microbalance. The second Au deposition step is performed at the angle $\theta_2=+30°$ with respect to the sample normal (positive angles correspond to a clockwise rotation), over the side of the ripple profile directly facing the ion beam during the nanopatterning step (Fig. 1d). Now the Au beam forms a 20° angle with respect to the "steep" facets and a 65° angle with respect to the "wide" facets local surface normal; the situation is now opposite compared to $\theta_1$ conditions and $h_{left-2}=h_0 \times \cos 65°$ while $h_{right-2}=h_0 \times \cos 20°$. To grow an approximately uniform Au layer over the asymmetric slope modulate ripple profile it's sufficient to impose a desired thickness $h$ and solve one of the following equations for $h_0$:

$$h_{left-1} + h_{left-2} = h_0 \times \cos 20° + h_0 \times \cos 65° = h$$

$$h_{right-1} + h_{right-2} = h_0 \times \cos 65° + h_0 \times \cos 20° = h$$

As shown in the Scanning Electron Microscopy (SEM) cross sectional image of the sample after metal deposition (Fig. 1e) one can clearly identify the presence of the Au film with thickness $h$=25 nm (Fig. 1d) which conformally follows the profile of the glass template. Such a rippled air/Au-thin-film/glass template can be considered as an IMI (insulator-metal-insulator) structure with slanted facets. We stress that such asymmetric, so-called slanted (or blazed), quasi-1D gratings cannot be fabricated by top-down techniques unless complex gray-scale lithography processes are adopted.

For the excitation of the propagating plasmon it is necessary that proper coupling of light momentum takes place at the metal/dielectric interface, mediated by transfer of the grating wavevector $k_G = n\frac{2\pi}{\Lambda}$ (see Refs. [5,6] and references therein). The latter process additionally requires that the spatial coherence of the grating pattern (Figure 1f) is comparable with the transverse coherence length $L_t = 0.16\frac{\lambda D}{\delta} = 9.6$ μm of the exciting thermal light beam at a nominal wavelength $\lambda = 600$ nm, collimated by an optical fiber with a core diameter $\delta = 1$ mm placed at a distance D= 0.1 m from the grating. Indeed the 2-dimensional self-correlation analysis of the sample AFM topography (see Fig. S3) demonstrates the visibility of coherent oscillations at the grating periodicity $\Lambda \approx 200$ nm even for lateral translations of the order of 10 μm which are comparable to the typical transverse coherence length $L_t$ of incoming photons. This striking degree of long range morphological order, considering the fully self-organized nature of our diffraction nanogratings, enables the possibility to investigate them as large-area launchers of SPP modes at optical frequencies.

Normal incidence linearly polarized optical transmission measurements are shown in Fig. 2a.

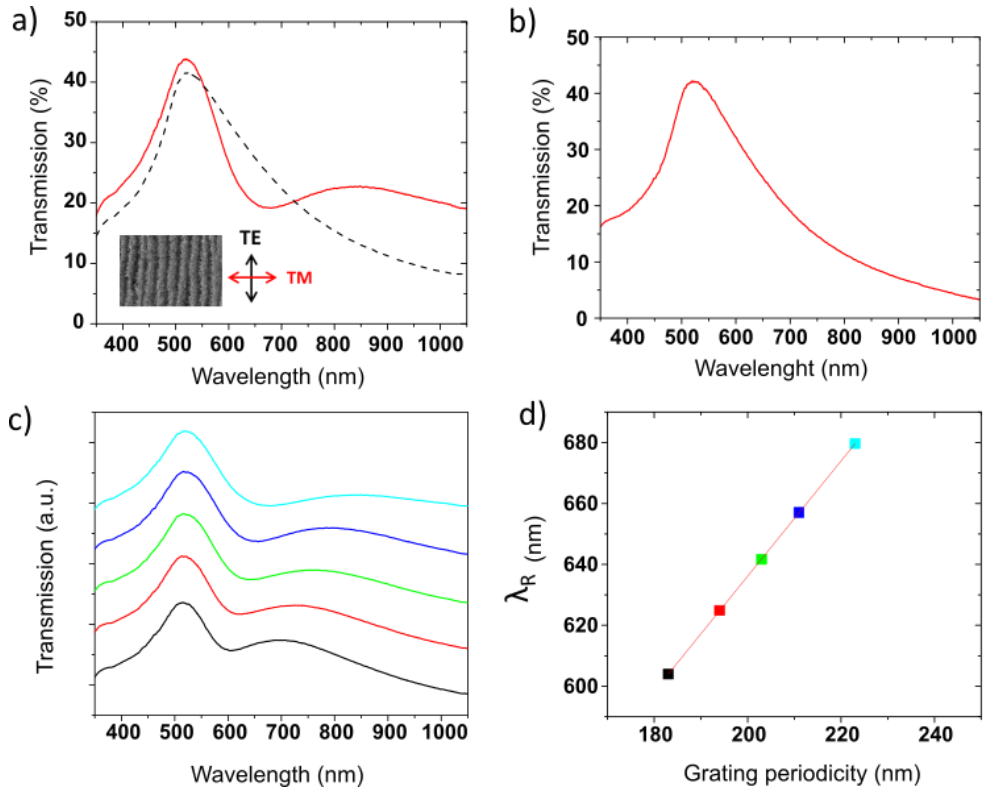

***Figure 2** – a) Normal incidence transmission measurements of a function of polarization. TM polarization corresponds to the continuous red curve, while TE polarization corresponds to the black dashed curve, as sketched in the panel inset. b) Polarization independent transmission spectrum of a flat 25 nm thick Au film on bare glass. c) TM normal incidence transmission spectra acquired for different periodicity of the Au/soda-lime blazed grating; 183 nm (black curve), 194 nm (red curve), 203 nm (green curve), 211 nm (blue curve), 223 nm (cyan curve). d) Plot of resonant wavelength vs. grating periodicity. All the transmission spectra shown in this panel are normalized to the one of a flat bare glass.*

For TE polarization (Fig. 2a, dotted black curve) the electric field oscillates parallel to the diffraction grating grooves (ripples long axis); in the considered spectral range, the sample transmission curve resembles the one of a Au flat thin film of equivalent thickness (Fig. 2b). For TM polarization (Fig. 2a – continuous red curve) the electric field oscillates orthogonally to the diffraction grating grooves (ripples short axis) and a strong dichroism is observed as transmission shows a broad minimum centered at the wavelength λ of about 660 nm. This transmission dip (corresponding to a maximum in the extinction spectrum) is attributed to the excitation of a propagating SPP mode along the continuous rippled Au film, which is enabled by the momentum exchange mediated by photon-grating interaction [21,23,33]. In Fig. 2c we demonstrate that the SPP resonant wavelength, measured in the transmission spectra under TM polarization at normal incidence, can be tuned by varying the grating periodicity at different samples coordinates along our linearly graded nanogratings. By analyzing the 2D self-correlation function of the AFM topographical images [44] we can associate the grating periodicity co-localized with the optical transmission spectra (see Fig. S2). In Fig. 2d the wavelength of the transmission dip minimum of each spectrum is plotted against the corresponding grating periodicity. The SPP resonant wavelength redshifts linearly and monotonically starting from the smaller considered periodicity (Λ=183 nm – black curve in Fig. 2c) to the bigger one (Λ=230 nm – cyan curve in Fig. 2c). Such a monotonic redshift of the SPP wavelength with increasing grating periodicity Λ is to be expected since a smaller momentum exchange $k_G = n\frac{2\pi}{\Lambda}$, mediated by photon-grating interaction, depends reciprocally from the period (n is an integer number either positive or negative) [5,6].

To further investigate and strengthen the attribution of the observed optical extinction loss to the excitation of an SPP mode, transmission measurements as a function of the sample tilt relative to the light beam were performed. The sample is illuminated from the glass side and the signal is coaxially collected in extinction configuration in TM polarization. The sample is tilted forming an angle θ between the incident light beam and the surface normal (Fig. 3a). The sample rotation is performed counter-clockwise, that is reducing the angle formed by the incident light beam and the local surface normal of the wide facets of the rippled Au/glass template (Fig. 3a). In Fig. 3b transmission spectra are plotted for different sample tilts, increasing in 10° steps from 0° to 60°.

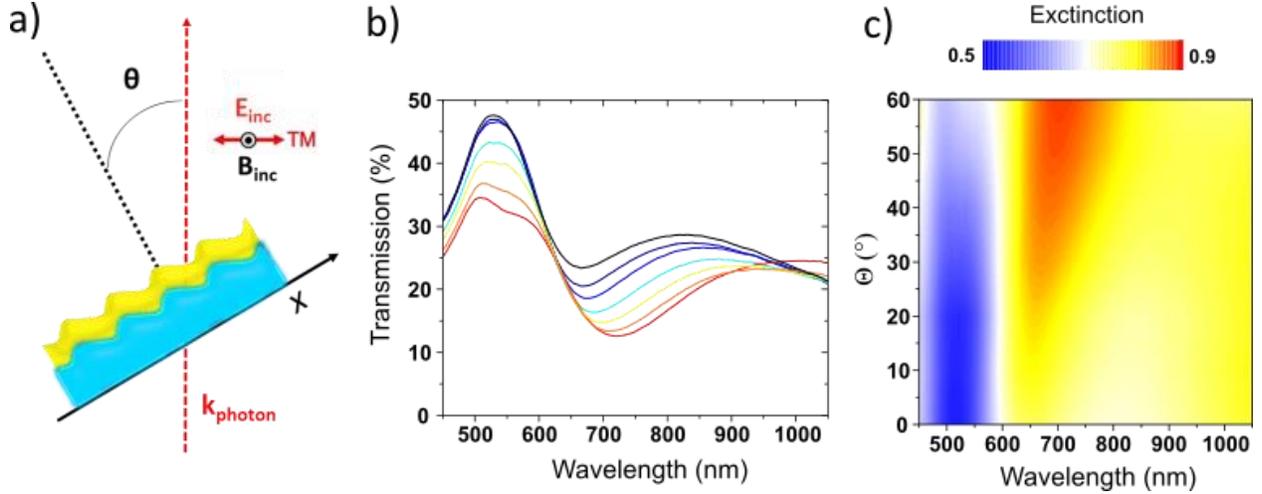

*Figure 3* – a) Sketch of the tilted transmission (T) measurements configuration. The red arrow $k_{photon}$ represents the light beam incoming from the glass side. Θ is the angle formed between the light beam and the normal to the flat sample surface. The x arrow defines the axis along the metal dielectric interfaces. In the inset the direction of oscillation of the incoming electric field $E_{inc}$ (red arrow) and the out of plane magnetic field $B_{inc}$ (black dot) are represented. b) TM transmission measurements as a function of Θ; 0° (black curve), 10° (royal blue curve), 20° (blue curve), 30° (cyan curve), 40° (yellow curve), 50° (orange curve), 60° (red curve). c) Extinction maps (calculated from the transmission spectra as $\frac{100-T(\%)}{100}$) as a function of both wavelength and tilt angle Θ.

The transmission dips redshift monotonically with increasing sample tilt, confirming the dispersive nature of the observed propagating SPP mode. For the highest tilt θ=60° (red curve in Fig. 3b) an additional, weak relative transmission minimum is observed at about λ=540 nm. In order to better visualize the experimental trend, in Fig. 3c the wavelength dependent extinction map (computed as $\frac{100-T(\%)}{100}$) is plotted against both wavelength λ and sample tilt θ. The non-dispersive blueish extinction band centered at about 530 nm is clearly associated with the spectral edge at the onset of Au interband transitions, while the strong and dispersive red band describes the SPP mode which intensifies and broadens with increasing sample tilt.

The monotonic redshift of the SPP resonance with increasing sample tilt (i.e. with increasing projection of photon wavevector $k_{ph}$ in the positive x-direction) suggests the unidirectional counter-propagating nature of the SPP mode with respect to the incident light beam. This trend can be reconciled considering that the grating allows to exchange momentum $k_G = n\frac{2\pi}{\Lambda}$ in discrete units and the effective wavevector component $k_x$ along the *x* axis can be derived according to the well-known relation: $k_x = |k_{ph}|\sin\theta + n\frac{2\pi}{\Lambda}$ where $k_{ph}$ is the modulus of photon wavevector, θ is the angle formed by the light beam with the normal to the flat sample surface (Fig. 3a), λ is the periodicity of the Au/soda-lime grating in the region of the optical spot and *n* is an integer number, either positive or negative. In particular, given the wavevector of the photon in correspondence to the SPP mode and the periodicity Λ of the grating, one observes that for negative n values $k_x$ is negative and decreases in modulus for increasing values of θ.

In order to better highlight the observed trend of the dispersion relation of the surface plasmon, in Figure 4a we plot jointly the experimental SPP energy $\hbar\omega_{SPP}$ and the theoretical value derived for a flat Au slab as a function of the in plane wavevector using the conventional expression $k_{SPP} = \frac{\omega_{SPP}}{c}\sqrt{\frac{\varepsilon_1\varepsilon_2}{\varepsilon_1+\varepsilon_2}}$ where $\varepsilon_1$ is the real part of the dielectric function of a thin flat Au film [52], and $\varepsilon_2$ is the real part of the effective dielectric function of the corrugated glass supporting the Au slab. The plotted SPP branch - light blue line - corresponds to the counter-propagation of the SPP confined at the Au-glass interface with effective dielectric refractive index $n_2$=2.3 (the straight dashed line is the dispersion relation of light in the dielectric medium). The experimental values of the SPP energy $\hbar\omega_{SPP}$, associated to the transmission dips of Figure 3b, are plotted in terms of $k_x = |k_{ph}|\sin\theta + n\frac{2\pi}{\Lambda}$ assuming n=-1, i.e. incoming photons with positive momentum $k_x$ couple with a SPP with negative $k_x$, via exchange of grating coupled negative momentum $k_{G,-1} = -\frac{2\pi}{\Lambda}$.

We stress that the comparison of Figure 4, though simplified and based on an effective medium approximation, allows to describe the general trend of SPP dispersion when the Au film is so thin (t=25 nm) that coupling between plasmon polaritons propagating at the two interfaces of the slab cannot be neglected. This leads to a hybridization of the SPP modes of a thick metal slab into an antisymmetric mode (out-of-phase matching of electric fields at the two sides of the metal film also called short-range hybrid mode), which is red-shifted compared to the single-interface metal/glass SPP, and into a symmetric mode (in-phase matching of electric fields at the two sides of the metal film also called long range hybrid mode), which is blue-shifted compared to the single-interface air/metal SPP [53–56]. The best match with the experimental data (Figure 3b,c) corresponding to the low energy antisymmetric mode is found for an effective refractive index of the corrugated glass/Au slab $n_2$=2.3, a considerably higher value compared to the typical n=1.5 of glass which would be expected in the thick Au film limit. In our experimental configuration, the high energy symmetric hybrid SPP mode is not observable since it is strongly damped by the Au s-d interband transitions which are active below 520 nm.

In Figure 4b, instead, we highlight the possibility to modify the experimental SPP plasmon energy $\hbar\omega_{SPP}$ by just varying the grating periodicity $\Lambda$ (see Fig. 2d), when excitation takes place at constant incidence angle (normal incidence, $\theta$=0°). The theoretical value of the SPP energy

$\hbar\omega_{SPP} = \hbar c\, k_{SPP} \sqrt{\frac{\varepsilon_1+\varepsilon_2}{\varepsilon_1\varepsilon_2}}$ evaluated for $k_{SPP} = k_{x,-1} = |k_{ph}|\sin\theta + n\frac{2\pi}{\Lambda}$ in correspondence to the diffracted order n= -1 and θ=0°. Matching between experimental and theoretical values, when the grating period progressively increases from 183 nm to 223 nm, is achieved by optimizing the only free parameter, the effective refractive index of the undulated dielectric layer $n_2$ supporting the Au film, in a narrow range (2.28±0.08).

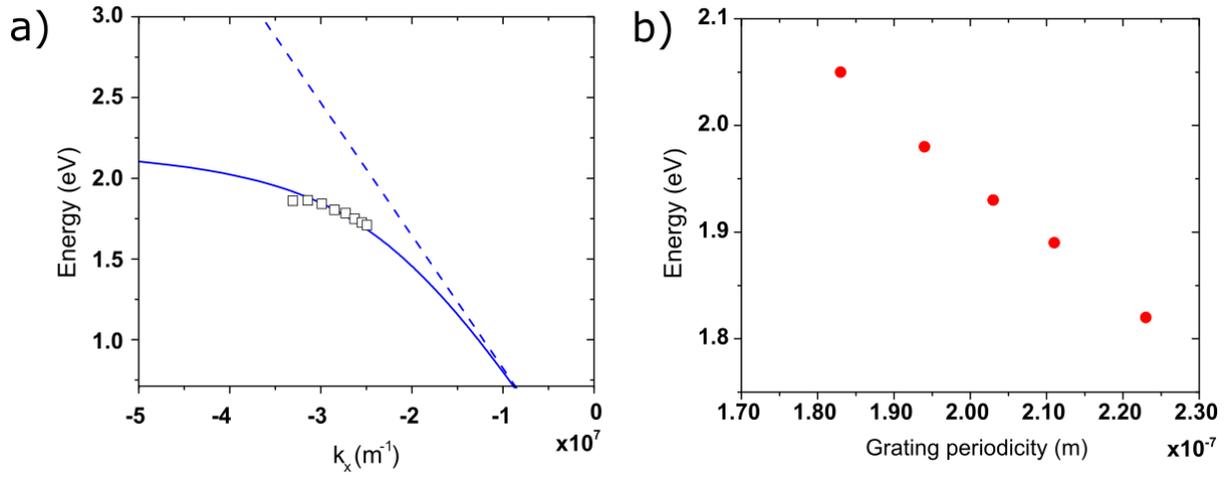

*Figure 4 – Dispersion relation of the short-range antisymmetric SPP mode $\hbar\omega_{SPP}$ vs. $k_x$ propagating at the Au/dielectric interface (solid blue curve) and light line of a photon travelling in the dielectric effective medium (dashed blue curve). Experimental data are plotted as squares. (b) Plot of resonant energy $\hbar\omega_{SPP}$ of the short-range hybrid SPP mode at normal incidence vs. the grating periodicity.*

Similar to the case of localized plasmon resonances supported by nanoparticles, the incoming electric field is strongly enhanced and confined near the dielectric interface of the metal film supporting an SPP. In the thin film regime explored in our experiment (Au thickness h=25nm), the evanescent field of the short-range hybrid SPP is able to reach the outer half layer occupied by air where it can couple with molecular probes there located [53–55]. With this in mind we recurred to SERS/micro-extinction measurements in order to probe the SPP near-field enhancement on the rippled Au/soda-lime blazed grating. Methylene blue (MB) was used as probe molecule (Methods, paragraph 2.4). Different sample positions a few mm apart from each other have been probed in order to verify the dependency of the SERS signal from the varying periodicities of our graded nanogratings (Fig. 2c). For each considered sample position, TM polarized SERS and co-localized optical micro-extinction measurements are performed (Methods, paragraph 2.4). As a reference, we acquired the Raman signal of MB absorbed under the same experimental conditions on a flat Au film with a thickness of 150 nm (the SERS signal - red line - is negligible when plotted in linear scale). This allows us to estimate the SERS gain of our rippled substrate, defined as the ratio between the intensity of the peak at 1625 cm$^{-1}$, measured on the rippled sample (e.g. black curves in Fig 5 a,b), and on the flat Au film (red curves in Fig. 5 a,b), normalized to the laser power and integration time [40]. The so defined SERS gain, plotted as a function of the grating periodicity in each considered position for the 638 and 785 nm pump lasers, shows a monotonic increasing trend as reported in Fig. 5c and Fig. 5d respectively.

As previously discussed, the SPP resonance spectrally shifts when moving at different sample coordinates, due to a spatial dependence of the SO grating periodicity (Fig. 2c). It's thus possible to nicely correlate the intensity of the plasmonic extinction strength at the pump lasers frequency, which is responsible for the SERS near field enhancement, with the co-localized measured SERS gains. After background subtraction the extinction values at the pump lasers wavelengths of 638 nm and 785 nm are then plotted as a function of different sample position in Fig. 5e and Fig. 5f, respectively (see Fig. S4 for details). Remarkably, the intensities of TM extinction and co-localized SERS gain are very strongly correlated for both pump laser wavelengths as a function of sample coordinates, thus confirming that the observed SERS signal is indeed matched with the strength of the SPP induced electric near field enhancement.

The SERS gain with the 638 nm laser turns out to be the order of $3 \times 10^2$. It must be noticed that at 638 nm an electronic transition in MB molecules is excited. In this case Surface Enhanced Resonant Raman (SERRS) excitation takes place which leads to a stronger enhancement of the Raman emission of MB. The signal amplification provided by the resonance effect (typically 2 orders of magnitude for MB), however, is expected to be independent from the substrate on which MB is deposited, so it is not supposed to influence our estimate of the SERS gain. On the other hand, for the pump laser at 785 nm, for which no resonant Raman excitation takes place, the maximum gain increases approaching the $10^3$ figure, which compares well with SERS substrates exploiting localized plasmon resonances of disconnected nanoparticles [40] and lithographically made gratings [38,39]. We can thus consider the present observations of relevance in view of molecular sensing applications since it is possible to reduce the NIR laser pump power (and in turn reduce sample damage and fluoresce background) while still keeping relevant the SERS gain in the near infrared range of the spectrum [57,58].

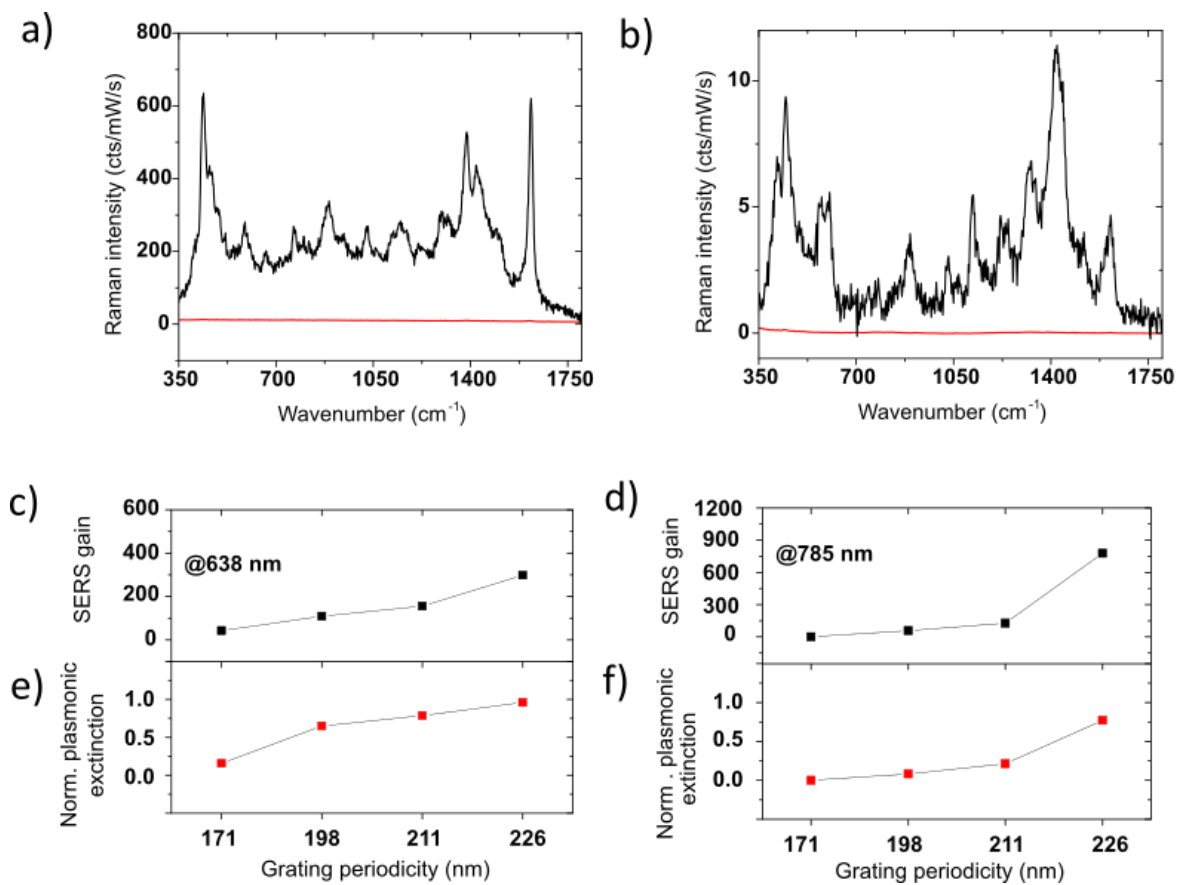

*Figure 5* – *a, b) Sample SERS spectra (black traces) and reference flat Au film Raman signal (red traces) for 638 nm and 785 nm pump lasers, respectively. c, d) SERS gain of MB molecules signal with 638 and 785 nm pump lasers, respectively, for different grating periodicity. e, f) Plasmonic extinction values at 638 and 785 nm, respectively, as a function of grating periodicity, derived from Raman co-localized transmission measurements (see Fig. S4).*

We stress that the discussed SERS gains could indeed be further increased by optimizing easily accessible experimental parameters. We can mention the increase of the grating period resulting in a redshift, beyond the onset of interband transitions, of the high energy long-range hybrid SPP, which is characterized by a longer lifetime and a transverse electric field penetrating more deeply into the dielectric environment where probe molecules are found [54]. Morevoer an optimization of the Au thin film thickness to find the ideal hybridization and photon-SPPs coupling conditions could further maximize the SERS gain. Alternatively, the plasmonic response of our linearly graded nanogratings could be further tuned in view of plasmon enhanced spectroscopies by e.g. matching the pump laser frequency across the UV to NIR spectral range by growing conductive layers with different free-carrier density. The SPP dispersion relation is in fact strongly affected by the metal plasma frequency [5,6]: a blueshift towards the UV could be achieved by growing on top of the rippled dielectric template metal films with high density of free carriers such as Ag and Al, while following a similar approach, one could imagine shifting the SPP excitation in the NIR-IR range by growing a thin film with a reduced free carrier density like e.g. semiconductors with variable doping and 2D materials [59,60]. The wide spectral tunability of the plasmonic self-organized nanogratings, combined with the possibility to easily scale-up the low-cost and large area fabrication method here described represent crucial features in view of a broad range of real word applications aiming at photon harvesting and at advanced molecular sensing.

## 4. CONCLUSIONS

In this work we demonstrate how a self-organized rippled soda-lime glass template, prepared by an innovative wrinkling-assisted Ion Beam Sputtering (IBS) method, can yield large area, sub-wavelength, quasi-1D blazed grating, enabling the unidirectional excitation of propagating hybrid Surface Plasmon Polaritons (SPP) modes at the interface of thin Au conductive film conformally grown on top of it. For this purpose, we engineered the IBS process to fabricate nanogratings with graded periodicity which act as tunable SPP couplers/launchers, thus facilitating the optimization of operating condition by e.g. matching the pump laser frequency in plasmon enhanced spectroscopies. A rippled Au/soda-lime grating sample showed remarkable Surface Enhanced Raman Scattering (SERS) activity in the detection of methylene blue molecules. We demonstrate how the SERS gain is strongly correlated with the electric field enhancement produced by the excitation of counter-propagating hybrid SPP modes by comparing SERS and co-localized micro-extinction measurements. The SERS gain observed employing a 785 nm pump laser is particularly striking, making the sample interesting for molecular and bio-sensing applications with low fluorescence background. Indeed, the SERS gain could be increased by further optimizing fabrication parameters such as the thin conductive film thickness and grating periodicity. The large area self-organized SPP launching platform is very versatile since, by simply choosing the metal to conformally grown of top of the glass rippled template, it's possible to tailor the plasmonic response in a wide spectral range extending from the near UV to the near IR, and possibly beyond considering the use of semiconducting and 2D materials.

**ASSOCIATED CONTENT**

Supporting information is available online.

**ACKNOWLEDGMENTS**

Financial support is gratefully acknowledged from Ministero dell'Università e della Ricerca Scientifica (MIUR) through the PRIN 2015 Grant 2015WTW7J3, from Compagnia di San Paolo in the framework of Project ID ROL 9361. F.B.d.M. thanks Roberto Chittofrati and Ennio Vigo for providing technical support.

Mortensen, N. A. Plasmon–Phonon Coupling in Large-Area Graphene Dot and Antidot Arrays Fabricated by Nanosphere Lithography. *Nano Letters* **2014**, *14* (5), 2907–2913. https://doi.org/10.1021/nl500948p.

# Supporting Information

Fig. S1a shows the 2D self-correlation function of the self-organized glass nanograting AFM topography presented in Fig. 1a of the main manuscript, computed by means of Gwyddion software. The grating average periodicity λ is calculated by measuring the real space distance between the maximum and the secondary neighboring peaks in the 2D self-correlation line profile of Fig. S1b.

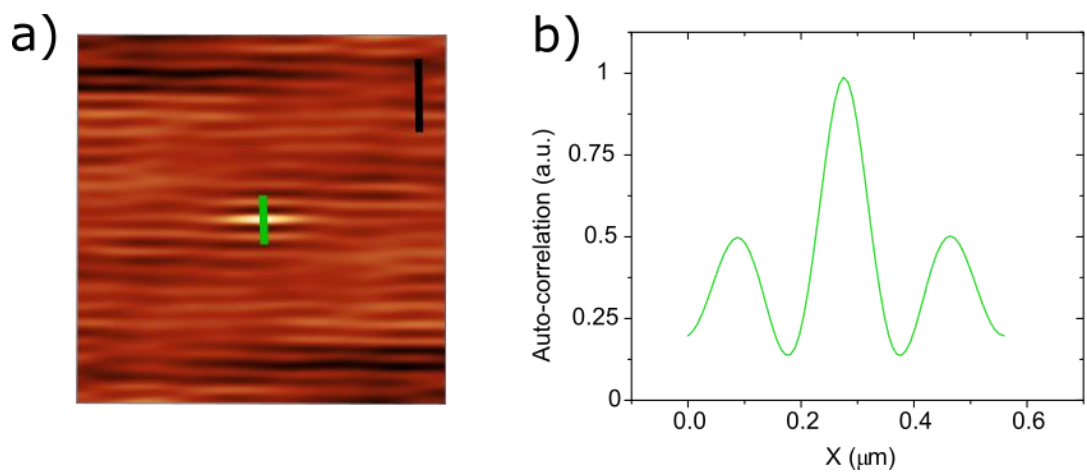

***Figure S1** – a) Self-correlation of the AFM topography shown in Fig. 1a of the main manuscript. The black scale bar corresponds to 800 nm. b) Self-correlation line profile corresponding to the green line in panel a).*

Fig. S2 shows the linear grating periodicity gradient as a function of sample position along the incident ion beam direction due to our engineered self-organized Ion Beam Sputtering fabrication process. The grating periodicitiy λ is measured by means of AMF microscopy through the 2D self-correlation of topographies as previously described.

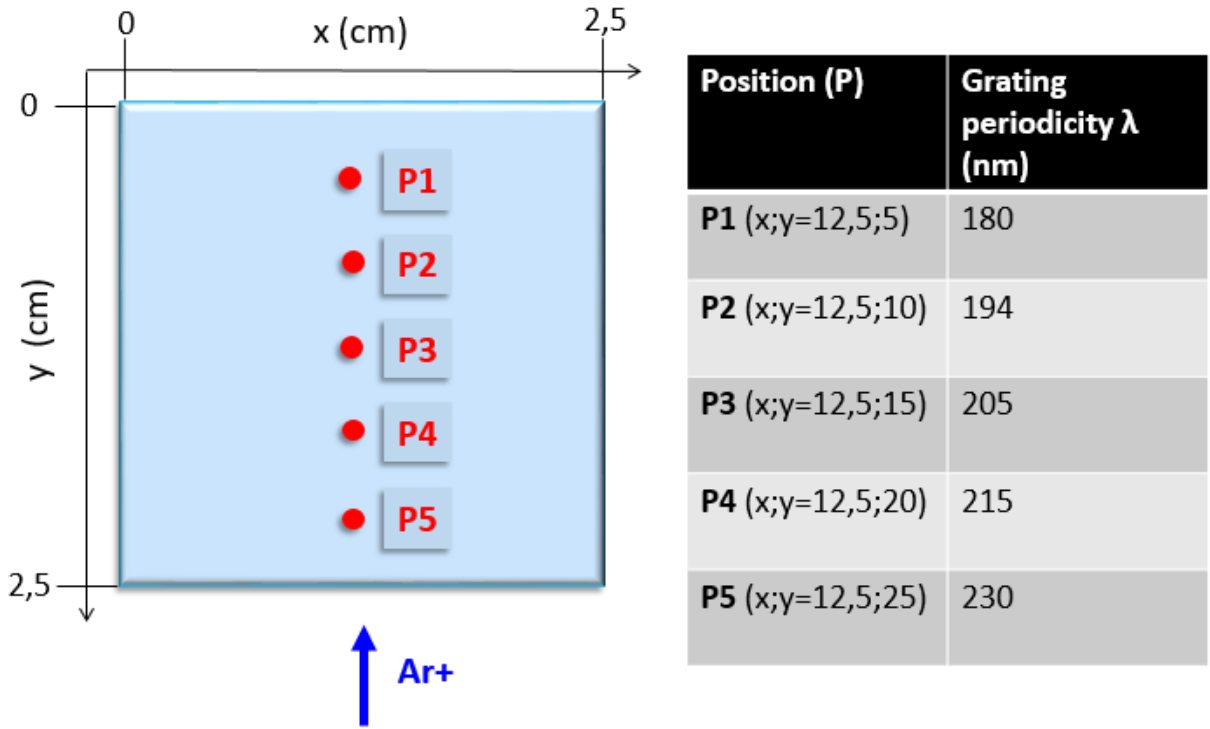

***Figure S2** – A sketch of the sample: red dots indicate the coordinates of different AFM measurements, while the blue arrows indicates the incoming Argon ion beam direction (left panel). The corresponding measured grating periodicity λ is reported in the table (right panel).*

In Fig. SI 3 we demonstrate the long-range morphological properties of the self-organized Au/glass nano-rippled grating, thought the analysis of a large area 2D auto-correlation extracted from AFM topography measurements (Fig. S3a). Remarkably, in the self-correlation line profile of Fig. S3c (corresponding to the green line in Fig. S3b), which has been measured by moving more than 10 µm away from the origin of the auto-correlation map, the modulations at the characteristic period around 200 nm persist well visible. This ensures that the grating morphological coherence can spatially match the incoming photons transverse coherence length in our experimental conditions, as required for SPP coupling to occur.

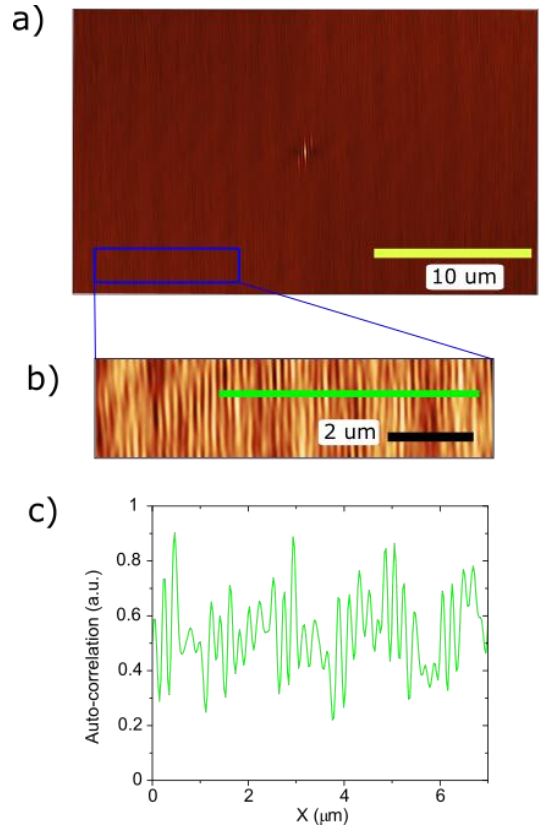

*Figure S3 – a) 2D auto-correlation of a AFM topography acquired on the Au/glass self-organized nanograting. b) Detail of panel a) highlighting a region several microns away from the central 2D auto-correlation maximum. c) Auto-correlation line profile corresponding to the green line in panel b).*

In Fig. S4a we plot the SERS co-localized transmission measurements for the different grating periodicities considered in Fig. 5 of the main manuscript. The measurements are normalized to the spectrum of a bare flat glass. Extinction is computed from transmission as 1-T and plotted in Fig. S4b after optical background is subtracted to isolate the plasmonic spectral fingerprint. Spectra are normalized to the most intense peak (yellow curve). The normalized plasmonic extinction values gain reported in Fig. 5e,f are extracted in correspondence of the pump laser frequencies, 638 and

785 nm respectively.

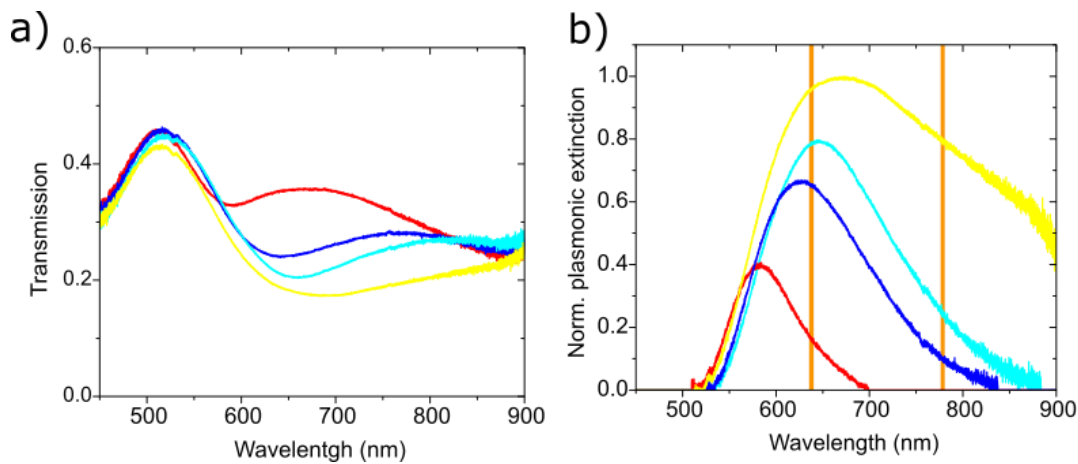

*Figure S4* – *a) Transmission spectra colocalized with SERS measurments acquired at different sample coordinates corresponding to the following grating periodicities: 171 nm (red curve), 198 nm (blue curve), 211 nm (cyan curve), 226 nm (yellow curve). b) Exctincion is computed from the transmission spectra of panel a) and the optical background is subtracted from the curves. Data is then normalized to the most intense peak. The two orange lines correspond to the wavelenghts of the two pump lasers, 638 and 785 nm respectively.*